%% file: main.tex
\documentclass[letterpaper, 10 pt, journal]{IEEETran}

\IEEEoverridecommandlockouts                              %

\usepackage{url}
\input{preamble}

\IEEEoverridecommandlockouts
\begin{document}

\title{Axes for Sociotechnical Inquiry in AI Research}
% \title{AI Development for the Public Interest: From Abstraction Traps to Sociotechnical Risks}
% sociotechnical lens; society lens in technical research
% Sociotechnical Stakes of AI Development

% \author{
% \thanks{Authors arranged alphabetically to indicate equal contribution.}
% % \IEEEauthorblockN{Anonymous for Submission}
% %\IEEEauthorblockN{McKane Andrus}
% %\IEEEauthorblockA{Partnership on AI \\
% %mckane@partnershiponai.org}
% \and
% \IEEEauthorblockN{Sarah Dean}
% \IEEEauthorblockA{Electrical Engineering and Computer Sciences\\ UC Berkeley \\
% dean\_sarah@berkeley.edu}
% \and
% \IEEEauthorblockN{Thomas Krendl Gilbert}
% \IEEEauthorblockA{Center for Human-Compatible AI \\ UC Berkeley \\
% tg340@berkeley.edu}
% \and
% \IEEEauthorblockN{Nathan Lambert}
% \IEEEauthorblockA{Electrical Engineering and Computer Sciences\\ UC Berkeley \\
% nol@berkeley.edu }
% \and
% \IEEEauthorblockN{Tom Zick}
% \IEEEauthorblockA{Berkman Klein Center for Internet and Society \\ Harvard University \\
% tzick@cyber.harvard.edu}
% }


\author{
Sarah Dean, 
Thomas Krendl Gilbert, 
Nathan Lambert and
Tom Zick %\thanks{Manuscript submitted on Feburary 13th, 2021.}
\thanks{First and third authors are with the Department of Electrical Engineering and Computer Sciences, University of California, Berkeley (e-mails: \texttt{ \{dean\_sarah, nol\}@berkeley.edu}). 
The second author is with the Center for Human-Compatible Artificial Intelligence, University of California, Berkeley (e-mail: \texttt{ tg340@berkeley.edu}). 
The final author is with the Berkman Klein Center for Internet and Society, Harvard University (e-mail: \texttt{tzick@cyber.harvard.edu}). }

% \IEEEauthorblockN{
% %McKane Andrus\IEEEauthorrefmark{1},
% Sarah Dean\IEEEauthorrefmark{1},
% Thomas Krendl Gilbert\IEEEauthorrefmark{2}, 
% Nathan Lambert\IEEEauthorrefmark{1} and
% Tom Zick\IEEEauthorrefmark{3}}
% \IEEEauthorblockA{\textit{Authors arranged alphabetically. }
% %\IEEEauthorrefmark{1}Partnership on AI, San Francisco, CA.\\
% % San Francisco, California\\ 
% % Email: mckane@partnershiponai.org
% }
% \IEEEauthorblockA{\IEEEauthorrefmark{1}Department of Electrical Engineering and Computer Sciences, 
% University of California, Berkeley. \\
% % Berkeley, California
% % Email: \{dean\_sarah, nol \}@berkeley.edu
% }
% \IEEEauthorblockA{\IEEEauthorrefmark{2}Center for Human-Compatible Artificial Intelligence, 
% University of California, Berkeley. \\
% % Berkeley, California
% % Email: tg340@berkeley.edu
% }
% \IEEEauthorblockA{\IEEEauthorrefmark{3}Berkman Klein Center for Internet and Society, Harvard University. \\
% % Emails: 
% \small{ %mckane@partnershiponai.org, 
% \{dean\_sarah, tg340, nol\}@berkeley.edu, tzick@cyber.harvard.edu}
% % Boston, Massachusetts,
% }

}

\newcommand{\shorttitle}[0]{TODO}

\markboth{IEEE Transactions on Technology and Society}
{Dean \MakeLowercase{\textit{et al.}}: \shorttitle}

\maketitle

\begin{abstract}
\input{0_abstract}
\end{abstract}


% \fi
% \IEEEpeerreviewmaketitle

\input{1_intro}

\input{2_types_trim}

\input{3_socioqs}

\input{4_case}

\input{5_concl}
\section*{Acknowledgments}
The authors would like to thank McKane Andrus for his comments and feedback on an earlier version of this paper and
the Center for Long-Term Cybersecurity for funding.
SD is supported by an NSF Graduate Research Fellowship under Grant No. DGE 1752814.
%\nol{Who are our grants}.

% We thank... TODO: ack funding



%\bibliographystyle{IEEEtran}
\bibliographystyle{IEEEtran}
\bibliography{ethics}  % .bib
% \input{_bios}

\end{document}

%% file: preamble.tex
\usepackage{booktabs}
\usepackage{graphicx} % for EPS use the graphics package instead
\usepackage{balance}  % for useful for balancing the last columns
\usepackage{booktabs} % for pretty table rules
\usepackage{ccicons}  % for Creative Commons citation icons
\usepackage{ragged2e} % for tighter hyphenation
\newcommand{\sect}[1]{Sec.~\ref{#1}}
\newcommand{\fig}[1]{Fig.~\ref{#1}}

\newcommand{\tab}[1]{Tab.~\ref{#1}}

\usepackage{xcolor}

%% file: 0_abstract.tex
The development of artificial intelligence (AI) technologies has far exceeded the investigation of their relationship with society.
Sociotechnical inquiry is needed to mitigate the harms of new technologies whose potential impacts remain poorly understood.
%\nol{I don't think that it is that they are poorly understood that is the problem, its the scale of their potential impact w.r.t. how well they are understood}
To date, subfields of AI research develop primarily individual views on their relationship with sociotechnics, while tools for external investigation, comparison, and cross-pollination are lacking.
In this paper, we propose four directions for inquiry into new and evolving areas of technological development: 
\textit{value}---what progress and direction does a field promote, 
\textit{optimization}---how the defined system within a problem formulation relates to broader dynamics, 
\textit{consensus}---how agreement is achieved and who is included in building it, 
and 
\textit{failure}---what methods are pursued when the problem specification is found wanting.
The paper provides a lexicon for sociotechnical inquiry and illustrates it through the example of consumer drone technology.

%% file: 1_intro.tex
\section{Introduction}
Recent years have seen increasing public awareness of the profound implications of artificial intelligence (AI) applications and large scale data collection. 
It is now common for both large tech companies and academic researchers to motivate their work on AI as interfacing with the ``public interest," matching external scrutiny with new technical approaches to making systems fair, secure, or provably beneficial. 
Meanwhile, scholars in Science and Technology Studies (STS) are working to equip system designers with toolboxes to help ground these motivations~\cite{selbst2019fairness}, promulgating the conversation on critical making~\cite{ratto2011, hamraie2018} and value-sensitive design~\cite{valuedesign} towards technical practitioners.

{Due in part to a gap between the pace and investment in advancing fundamental AI technologies and reflecting on its potential harms,
identifying points of engagement between critical theory and various AI domains remains challenging.
Furthermore, emerging subdomains such as AI Safety, Fair Machine Learning (Fair ML), and Human-in-the-Loop (HIL) Autonomy
lack a common origin, and subsequently hold different conceptualizations about the relationship between their systems and society~\cite{andrus2021}. }
% Despite these efforts, identifying points of engagement between critical theory and various AI domains remains challenging, especially in emerging subdomains such as AI Safety, Fair Machine Learning (Fair ML), and Human-in-the-Loop (HIL) Autonomy. \nol{I think this sentance could transition better, I added a draft talking about the relative investment of technology itself over reflection, which could be nice}
% This is in part due to the lack of common origin, and subsequently different conceptualizations each subfield holds about the relationship between their systems and society~\cite{andrus2021}. 
While these differences make it challenging to develop interventions with field wide relevance, the distinct lines of inquiry pursued by each subfield present opportunities to research AI's increasingly central role in social relations. 
As sites of investigation, these domains are not only some of the most promising case studies for refining theories of sociotechnics over the next decades, but are also viable as beachheads for the normative appraisal of AI as a sociotechnical phenomenon.

To lay the groundwork for evaluating the relationship between AI and the sociotechnical, we portray these subfields as modes of sociotechnical inquiry in their own right. 
We do not forward a specific sociotechnical lens, ontology, or mode of critical reasoning in order to evaluate them. 
Instead, we frame these subfields' encounters with sociotechnics as \textit{the problem of defining the interface between systems and reality}. 
This problem requires the designer to manage how specific technical possibilities depart from and intervene on the state of the world \cite{rosa2020uncontrollability}, and to define (often implicitly) what success and failure look like in terms of design practice. 
These design practices may reveal uncertainties, alternate technical formulations, and counterfactual scenarios -- in the process generating unprecedented positions of sociotechnical engagement. 
From these positions, various interpretations of the design situation are simultaneously possible and amenable to inquiry. 
In this way the social---the matrix of relationships through which design interpretations are forged and considered---emerges \textit{sui generis} from the technical. 

\begin{figure}[t]
    \centering
    \includegraphics[width=0.98\columnwidth]{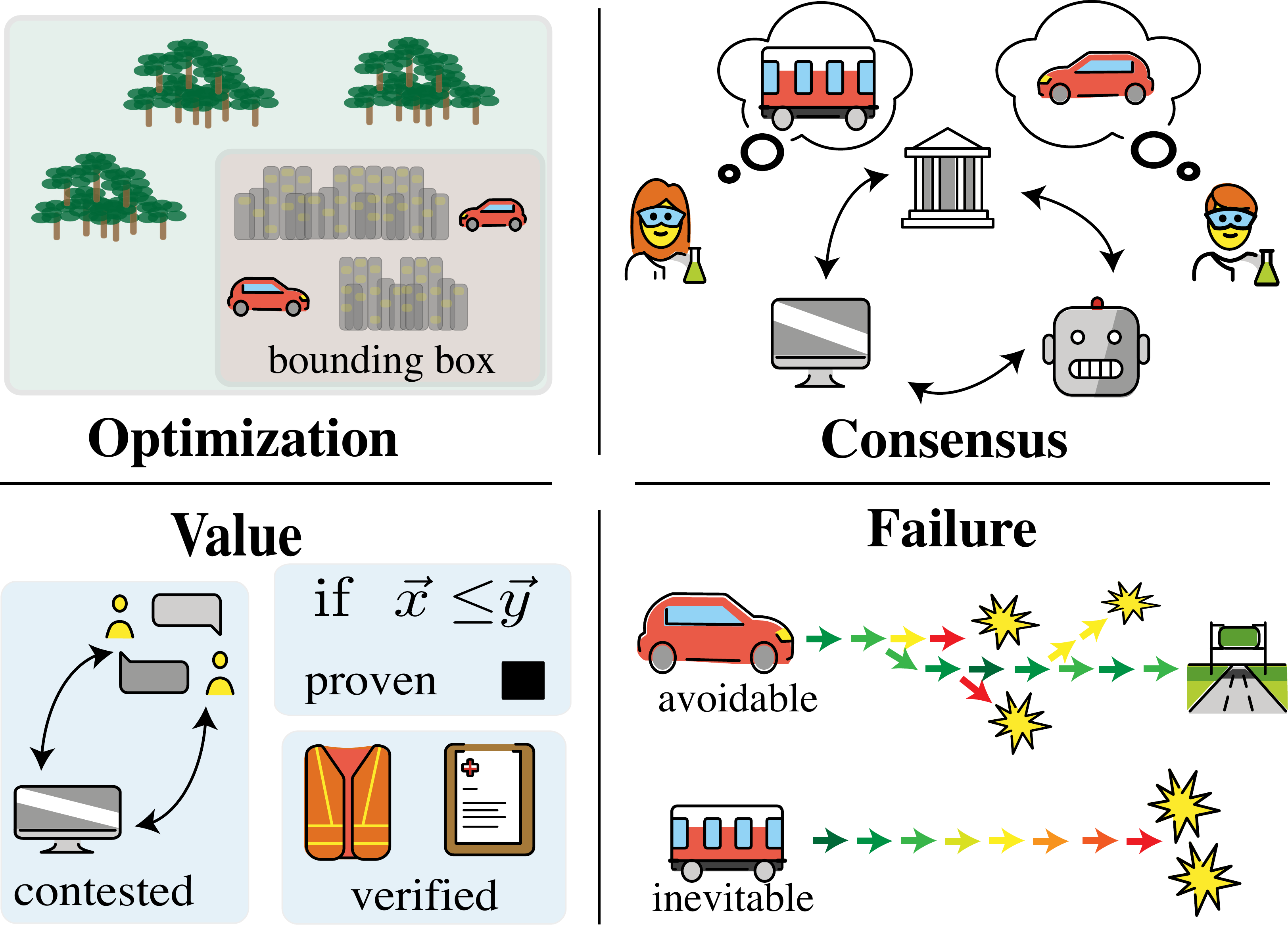}
    \caption{We propose four vectors for investigating the sociotechnical engagement of various branches of AI research: 
    \textit{value}---what does a field deem of merit, 
    \textit{optimization}---how the designed system within a problem formulation relates to the real-world dynamics, 
    \textit{consensus}---who and what is in the conversation towards agreement, 
    and \textit{failure}---what are the assurances if the system proves inadequate.
    Engaging with these axes is necessary to evaluate the potential social impact of emerging AI technologies.
    }
    \label{fig:concept}
\end{figure}

AI Safety, Fair ML, and HIL Autonomy pose the directions of inquiry differently to themselves, taking the form of distinct \textit{abstractions} whose assumptions mediate the divide between systems and social order. 
These abstractions entail different foci and procedures when technical problems arise, giving each community a distinctive research culture.
They have distinctive flavors of inquiry and values that guide that inquiry, particular spaces of investigation, and different limits to what one may or may not think about as ``in scope" (i.e. is able to be modeled). 
% While one field's parameters of inquiry often have substantial overlap with other subfields', when actualized the small differences can yield relationships with sociotechnics that vary widely.
{While the fields' parameters of inquiry often have substantial overlap, the small differences can yield relationships with sociotechnics that vary widely when actualized.}
The sociotechnical constraints on inquiry are thus irreducibly phenomenological: a space of interactive and affective concern that is shaped by researchers and in turn shapes the engineering and data science curricula that look to these subfields for guidance. 
Each space is not a set of precise definitions ready-made for computation, but a continuum whose elements are contested, and in which definitions are proposed, debated, refined, adopted, or rejected.

Textbooks, advisors, journals, and other socio-historical determinants pattern each space of concern and help practitioners internalize it. 
Still, as the technical possibilities expand in scope and significance, each community must come to terms with its own continuum and create new tools that purport to solve its own sociotechnical problems. 
This makes phenomenological comparison appropriate for revealing the distinctive elements of each space. 
Beyond historical genealogy, comparison helps us appreciate the tools for what they are: adjuncts to inquiry that help resolve or defer computational indeterminacies and semantic equivocations.

In this work we aim to reach two audiences. 
First, we map techniques of deliberation available to practitioners for the purpose of reflecting on abstractions in their own research. 
Second, we identify contested spaces of normative engagement for external scholars to call their own. 
We begin by reviewing the three subfields, pointing out key research themes. 
Next we present our axes of phenomenological comparison, namely \textit{value}, \textit{optimization}, \textit{consensus}, and \textit{failure} as highlighted in \fig{fig:concept}. 
We then conclude with a case study to illustrate the practical stakes of these differences of mindset in the context of a specific near-future AI system: consumer drone technology.
Together, our hope is that these audiences will make sustained and %substantive 
constructive engagement on sociotechnical questions achievable for other subfields that are still nascent or have not yet been created.

%% file: 2_types_trim.tex
\section{Historical Background}
\label{sec:subfields}

Recent examples of technical work grappling with the societal implications of AI include developing provably safe and beneficial artificial intelligence (AI Safety), mitigating classification harms for vulnerable populations through fair machine learning (Fair ML), and designing resilient autonomy in robotics and cyber-physical systems (HIL Autonomy). 
The methods by which the subfields define and pursue their own sociotechnical questions reflect upon the history and culture of the communities. 
In what follows, we outline the motivating concerns of each subfield and identify key developments in their evolution. For a more complete analysis see~\cite{andrus2021}.

\subsection{AI Safety}
The field of artificial intelligence (AI) has often situated itself within a wider disciplinary context. 
Famously, at foundational summits at Dartmouth and MIT, computer scientists, logicians, and psychologists came together to chart a course artificial intelligence researchers could follow 
to arrive at human-like capabilities with machines~\cite{russell2002artificial}.  
That course, which laid the groundwork for ``symbolic" or 
``good-old-fashioned" AI, had an incredible number of hiccups and was eventually overwritten. 
A similarly diverse group of disciplinary representatives moved the field away from symbolic and comprehensive logical reasoning to either more situated, interactionist understandings of cognition~\cite{dreyfus1992computers,agre1997computation} or to biologically-inspired, connectionist strategies of learning~\cite{sun2014connectionism}. 
This branch of AI has also borrowed much inspiration from the economic principles of \textit{mechanism design}, an approach to game theory in which the designer first specifies a desired outcome and then designs agent incentives so that the outcome is achieved~\cite{hadfield2016cooperative,leike2017ai,natarajan2010multi,kuleshov2015inverse,brundage2020toward}.

AI Safety began to solidify from these interdisciplinary underpinnings around growing 
%Following these interdisciplinary developments, there has been a 
concerns that AI systems could become capable of endangering humans and society writ large~\cite{bostrom2003ethical,yudkowsky2008artificial,armstrong2016racing}. 
Motivated both by longstanding concerns about the possibility of a ``technological singularity"~\cite{kurzweil2005singularity} as well as recent expansive applications of machine learning in critical infrastructure domains, many AI Safety promoters fear that AI research projects are approaching a level of capability that will expand beyond their control~\cite{kurzweil2005singularity, bostrom2017superintelligence}. 

Regardless of the likelihood of such a scenario, the nascent field of AI Safety has arisen to preemptively confront these dangers. 
% Despite recent high-profile endorsements from computer scientists and philosophers such as Stuart Russell \cite{russell2019human} and Nick Bostrom \cite{bostrom2017superintelligence}, given debates on when and if these concerns will manifest, AI Safety is still a nascent research community. 
{Despite recent high-profile endorsements from computer scientists and philosophers such as Stuart Russell \cite{russell2019human} and Nick Bostrom \cite{bostrom2017superintelligence}, AI Safety is still a nascent research community facing debates over when and if their concerns will manifest. }
At present there is no independent conference for this field, although workshops and panels on AI Safety have become a regular fixture of larger AI venues such as NeurIPS, ICML, and AAAI, while specific AI-safety oriented research labs (e.g. CHAI, OpenAI) host invited technical presentations on a semiweekly or monthly basis. 
The field has also attracted interest from research centers and philanthropic organizations dedicated to the study and mitigation of long-term existential risk, as well as industry leaders in AI.

%%%%%%%%%%%%%%%%%%%%%%%%%%%%%%%%%%%%%%%%%%%%%%%%%%%%%%%%%%%%%%%%%%%%%%%%%%%%%%%%%%%%%%%%%%%%%%%%%%%%%%%%%%%%%%%%%%%%%%%%%%%%%%%%%%%%%%%%%%%%%%%%%%%%%%%%%%%%%%%%%%

\subsection{Fair Machine Learning} % in Data-driven Systems}
The field of machine learning (ML)---primarily concerned with the science of making decisions and models from data---emerged in the late 1950s with the design of a self-improving program for playing checkers~\cite{samuel1959some} and quickly found success with static tasks in {pattern classification}, including applications like handwriting recognition~\cite{jnilsson1965learning}.
ML techniques work by detecting and exploiting statistical correlations in data, towards increasing some measure of performance.
A prominent early machine learning algorithm was the perceptron~\cite{rosenblatt1957perceptron}, an example of supervised classification, perhaps the most prevalent form of ML.
For example, a classifier (or model) trained with labelled examples will be evaluated by its accuracy in labelling new instances.
 The perceptron spurred the development of deep learning techniques mid-century~\cite{olazaran1996sociological}; however, they soon fell out of favor, only having great success in recent decades in the form of neural networks via the increasing availability of computation and data. 
 Many ML algorithms require large datasets for good performance, tying the field closely with ``big data.''

However, optimizing predictive accuracy does not generally ensure beneficial outcomes when predictions are used to make decisions, a problem that becomes stark when individuals are harmed by the classification of an ML system. 
%The \textit{inequality} resulting from system classifications is the central sociotechnical risk of concern to practitioners in this subfield of Fair ML. \sarah{I would remove previous sentence}
A growing awareness of the possibility for bias in data-driven systems developed over the past fifteen years, starting in the data mining community~\cite{pedreshi2008discrimination} and echoing older concerns of bias in computer systems~\cite{friedman1996bias}.

%\sarah{added this para back to give more historical context}
Interest in ensuring ``fairness'' was
further catalyzed by high profile civil society investigation (e.g. ProPublica's Machine Bias study, which highlighted racial inequalities in the use of ML in pretrial detention) and legal arguments that such systems could violate anti-discrimination law~\cite{barocas2016big}. 
At the same time, researchers began to investigate model ``explainability'' in light of procedural concerns around the black box nature of deep neural networks.
The research community around Fairness in ML began to crystallize with the ICML workshop on Fairness, Accountability, and Transparency in ML (FAT/ML), and has since grown into the ACM conference on Fairness, Accountability, and Transparency (FAccT) established in 2017.

By shifting the focus to fairness properties of learned models, Fair ML adjusts the framing of the ML pipeline away from a single metric of performance.
There are broadly two approaches: individual fairness, which is concerned with similar people receiving similar treatment~\cite{dwork_fairness_2012}, and group fairness which focuses on group parity in acceptance or error rates~\cite{barocas-hardt-narayanan}.
The details of defining and choosing among these \emph{fairness criteria} amount to normative judgements about which biases must be mitigated, with some criteria being impossible to satisfy simultaneously.
Much technical work in this area focuses on algorithmic methods for achieving fairness criteria through either pre-processing on the input data~\cite{calmon2017optimized}, in-processing on the model parameters during training~\cite{zafar2019fairness}, or post-processing on model outputs~\cite{hardt2016equality}.
%The community also is concerned with making models scrutable, towards goals of transparency and accountability, ensuring a type of sociotechnical openness.
% \sarah{we may want to add back a sentence about transparency (and maybe one about refusal)}
%The Fair ML community is oriented towards the sociotechnical, engaging actively with critiques from STS perspectives.

%\sarah{not sure if we should include any of the following points; some might be redundant with future sections but we should check}
Importantly, FAccT is a strong locus of interdisciplinary thought within computer science. Building upon model-focused concepts like explainability, blendings of technical and legal concepts of recourse~\cite{ustun2019actionable} and contestability~\cite{mulligan2019shaping} are employed to approach community goals of accountability. 
Similarly, there have been multiple calls to re-center 
%the 
stakeholders
%for which a system is made ``explainable" in order 
to understand how explanations are interpreted and if they are even serving their intended purpose \cite{miller2019explanation,bhatt2020machine}.

%%%%%%%%%%%%%%%%%%%%%%%%%%%%%%%%%%%%%%%%%%%%%%%%%%%%%%%%%%%%%%%%%%%%%%%%%%%%%%%%%%%%%%%%%%%%%%%%%%%%%%%%%%%%%%%%%%%%%%%%%%%%%%%%%%%%%%%%%%%%%%%%%%%%%%%%%%%%%%%%%%
\subsection{Human-in-the-Loop Autonomy}
As many of the earliest robotic systems were remotely operated by technicians, the field of robotics has always had visions of human-robot interaction (HRI) at its core~\cite{goodrich2008human}.
Early work was closely related to the study of human factors, an interdisciplinary endeavor drawing on engineering psychology, ergonomics, and accident analysis~\cite{bainbridge1983ironies}.
With advancements in robotic capabilities and increasing prevalence of autonomy, the interaction paradigm grew beyond just teleoperation to \emph{supervisory control}.
HRI emerged as a distinct multidisciplinary field in the 1990s with the establishment of the Institute of Electrical and Electronics Engineers (IEEE) International Symposium on Robot \& Human Interactive Communication.
By incorporating principles from the social sciences and cognitive psychology, HRI practitioners formulate a control problem with predictions and models of human behavior. 

%Now, 
Since then, digital technology has advanced to the point that many systems are endowed with autonomy beyond the traditional notion of a robotic agent, including traffic signal networks at the power grid. 
We thus consider the subfield of \emph{HIL Autonomy} to be the cutting edge research that incorporates human behaviors into robotics and cyber-physical systems.
This subfield proceeds in two directions: 1)  innovations in physical interactions via sensing and behavior prediction; 2) designing for system resiliency in the context of complicated or unstable environments. 
These boundaries are blurring in the face of novel computational methods for autonomy and the prospective market penetration of new technologies.

The emerging subfield of HIL Autonomy uses ideas from classical \textit{control theory} while trying to quantify and capture the risk and uncertainty of working with humans~\cite{baheti2011cyber,banerjee2011ensuring}. 
It thus inherits some of the culture around verifying safety and robustness through a combination of mathematical tools and physical redundancy due to a history of safety-critical applications in domains like aerospace.
Technical work in this area typically entails including the human as part of an under-actuated dynamical system \cite{sadigh2017active, wu2018stabilizing}, such as a un-modeled disturbance. 
Through this lens, human-induced uncertainty is mitigated by predicting behavior in a structured manner, maintaining the safety of the system through mathematical robustness guarantees such as an explicit \textit{reachability} safety criterion~\cite{bajcsy2019scalable}.

The extent to which HIL Autonomy engages with the sociotechnical is thus far limited. 
Human-centered research focuses on localized one-to-one interactions, while research considering more global interactions remains largely in the realm of the technical.
However, the critical ``alt.HRI" track at the Association for Computing Machinery (ACM) or IEEE Conferences on Human-Robot Interaction indicates an emerging interest in how robotic systems interact with society more broadly.
While our identification of this emerging subfield is perhaps more speculative than the previous two, the physical realization of AI technologies will remain a crucial site of sociotechnical inquiry.

%% file: 3_socioqs.tex
\section{Proposed Tools for Sociotechnical Inquiry}
In this section we outline axes that can be used to unpack the assumptions and bounding boxes %portray the phenomenology that 
used by different subfields to interpret the interface between their technology and society.  
These axes are intended to knit together a phenomenological articulation of the sociotechnical dimensions of AI and provide points of engagement for qualitative reflection. 
They can be understood in terms of four modalities of abstraction available to present and future AI systems: 
\textit{value}, different interpretations of the interface that delimit how abstractions are evaluated; 
\textit{optimization}, to what extent the system design in a problem formulation matches the true domain dynamics; 
\textit{consensus}, methods of reaching agreement on evaluations; 
\textit{failure}, how to handle and compensate for the limits of the abstraction and solution.
Key terms from this analysis with respect to the subfields of AI presented in \sect{sec:subfields} are shown in \tab{tab:tomstable}.
As each community is quite young, these tastes are not set in stone, and researchers do have the ability to cross-pollinate abstraction strategies between camps.

\begin{table*}[t]
    \centering
    % \begin{tabular}{p{11.5mm} p{24mm} p{22mm}  p{17.5mm}}
    \begin{tabular}{c l l  l}
        \toprule
        Inquiry & AI Safety & Fair ML & HIL Autonomy \\
        \midrule
        Value & Epistemic Alignment & Distributive Guarantees & Precise Control\\ %Precise Control \\ 
        \rule{0pt}{3mm} Optimization & Mechanism Design & Classification & Reachability \\
        \rule{0pt}{3mm} Consensus & Forecasting & Debate & Certification \\
        \rule{0pt}{3mm} Failure & Provable Benefit & Contestability & Stability \\ 
        \bottomrule \\
    \end{tabular}
    \caption{Axes of Comparison for Sociotechnical Inquiry. 
    These terms serve to highlight essential properties of each subfield along the set of axes we propose. 
    We elaborate and explore related properties in Section 3.}
    \label{tab:tomstable}
\end{table*}

% \begin{table}[t]
%     \centering
%     \begin{tabular}{r p{19mm} p{14mm}  p{21mm}}
%         \toprule
%         Inquiry & AI Safety & Fair ML & HIL Autonomy \\
%         \midrule
%         Value & Epistemic \newline Alignment & Distributive Guarantees & Human Control \\ 
%         \rule{0pt}{3mm}    Optimization & Mechanism \newline Design & Classification & Reachability \\
%         \rule{0pt}{3mm}Consensus & Forecasting & Debate & Certification \\
%         \rule{0pt}{3mm} Failure & Provable Benefit & Contestability & Resiliency \\ 
%         \bottomrule \\
%     \end{tabular}
%     \caption{Axes of Comparison for Sociotechnical Inquiry.
%     }
%     \vspace{-10pt}
%     \label{tab:tomstable}
% \end{table}

\input{3a_value}

\input{3b_opt}
\input{3c_consensus}

\input{3d_failure}

%% file: 3a_value.tex
\subsection{Value: What are the nature and metrics of value?}
The names AI “Safety”, “Fair” ML, and HIL “Autonomy” ought not be taken too seriously. 
There is broad agreement across communities that socially responsible and effective AI systems must be simultaneously safe, fair, and independent from constant human oversight~\cite{russell2015research,whittaker2018ai,shi2014fairness,alur2015principles}. 
Instead, these fields' differences are rooted in incommensurate interpretations of these labels: in terms of how they pertain to system design and which technical abstractions they largely take for granted. 
As such, these terms are better conceived as metaphors that more or less problematically outline what the subfields themselves see as their distinctive research contribution to the AI landscape, and how they claim to identify the key features of a well-defined system.

Each subfield has a particular style to its inquiry. 
This manifests in two ways: a fundamentally shared sense of the value 
in particular modes of abstraction, and an accepted range of play for elaborating, modifying, refining, or discarding particular features of the abstraction. 
Together, these elements constitute the distinctive “tastes” present in each community. 

AI Safety evaluates a given abstraction strategy in terms of its promise for \textit{epistemic alignment}, conveyed in answer to the following query adapted from~\cite{russell2019human}: 
\textit{Have all sources of human concern been modeled in terms of uncertainty, with probabilities able to be learned for every logically possible human preference?} 
The elegance of this approach is that it sidesteps a potentially fraught appraisal of what human values are at stake in a given domain, in favor of computationally modeling them by observing the human behaviors present in that domain. 
A well-designed system will thereby learn how to determine the right thing to do, and a designers’ job is merely to clear this threshold rather than purport to answer substantive questions of meaning or significance.

Yet there is some disagreement on whether and how designers can maintain this criterion. 
One example debate is the tension between alignment and corrigibility, i.e. the ability to intervene on and correct the behaviors 
learned by the system once it has been designed~\cite{soares2015corrigibility,carey2018incorrigibility}. 
Corrigibility would not be a pressing concern for a perfectly-designed system, as guarantees would be in place on exhaustive value learning and the perfect simulation of human preferences. 
At some limit, it is not even clear such a system would need to be overseen by humans at all. 
There is also tension between empiricist and theoretical approaches to epistemic alignment: is it better to model probabilities through unprecedented scales of computation and data collection~\cite{irving2019ai,yampolskiy2019personal}, or by refining the application of expected utility theory to particular domains~\cite{bostrom2012superintelligent,drexler2019reframing}? 

Fair ML instead focuses on the \textit{distributive guarantees} made possible by the abstraction~\cite{liu2018delayed}: 
\textit{Has a discursive formulation of distributive choices been framed or made tractable in relation to a contemporary computational procedure, whose implications are well-defined?} 
This field does not claim to translate human behaviors into abstracted forms of knowledge, but to translate the normative criteria already relied on by human institutions into an algorithmic setting. 
Because many of these criteria are implicit or lack specification, 
this translation constitutes a major research challenge.
It is bolstered by efforts to make the resulting abstractions transparent and accountable in addition to technically fair.

Values within Fair ML are partially articulated through contention regarding what transparency and accountability actually mean. 
One example is research on explainability, which tries to transform the system from a “black box” (i.e. a system whose outputs do not have a understandable cause) into one whose input-output mappings are paired with a descriptor that can be interpreted~\cite{samek2017explainable,rudin2019stop}. 
% Another inquiry is critiques of exclusively formal work insensitive to social context that are claimed to infect the culture of Fair ML~\cite{keyes2019mulching,lipton2018troubling}. 
{Critiques of exclusively formal work claim that the ``mathiness'' culture of the Fair ML community causes an insensitivity to social context~\cite{keyes2019mulching,lipton2018troubling}.}
This is perceived as problematic because the integrity of Fair ML research depends on leveraging computational tools to meaningfully represent human concerns, based on inquiry into the boundary between computational and discursive approaches to social problems. 
This tension is integral to the distinctive Fair ML research agenda~\cite{corbett2018measure,bennett2019point}. 

Finally, HIL Autonomy concentrates its attention on \textit{precise control} in the context of system operation~\cite{fisac2018general}: 
\textit{Has the system been set up to operate well in the presence of human activities, such that the boundary between human and system can be known and effectively controlled?} 
There is less of a focus on translating the lived experience of some human domain into an algorithmic model. 
Instead, there is a sustained effort to balance abstract learning procedures with domain-specific assumptions, in order to generate guaranteed means of control over system behavior. 
The designer must appraise the domain in terms of how it actually works, rather than via desired behaviors or distributive priorities.

In lieu of normative debates, there are outstanding questions related to robust learning and comparing various methods for human control. 
%A prominent example is reachability, which examines how to ensure that the system can reach states that are known to be safe or trustworthy against other, riskier alternatives \cite{lygeros1999controllers} even while learning a model of its own and the environment's physical capabilities \cite{zacharias2007capturing}. %\sarah{I would rephrase this as ``ensure that the system can reach states that are known to be safe or trustworthy against other, riskier alternatives \cite{lygeros1999controllers} even while learning a model of its own and the environment's physical capabilities \cite{zacharias2007capturing}''}
%Another is 
A prominent example is legibility as a criterion for determining the boundary between system and environment: autonomous vehicles may perform suboptimal behaviors (like acting submissively at four-way stops) if it aids coordination with human drivers \cite{dragan2013legibility,lichtenthaler2016legibility}. 
This is distinct from both transparency and corrigibility, as the goal is not to aid diagnostics or intervene on value learning but to incorporate legibility directly into the system behavior and leverage this to improve model performance.

%% file: 3b_opt.tex
\subsection{Optimization: What is the relationship between the system and the broader environment dynamics?}
The formulation of structured optimization problems is the primary avenue for the representation of value in technical fields.
% The primary representation of value of a field is through the abstraction it raises to formulate structured optimization problems. 
Optimization refers to how system performance is rendered effective or sufficient according to specified metrics. 
This reflects how domain dynamics figure into the model specification---what a system is able to do, how it is framed, and who can control or oversee its functioning. 
% The problem of optimization thus constitutes a separate axis of comparison between fields, which must decide which environmental dynamics are in or out of scope for technical work. 
{The problem of optimization thus constitutes a separate axis of comparison between fields, to examine the decision of which environmental dynamics are in or out of scope for technical work. }

In AI Safety, the dynamics of the system are not explicitly modeled, but are generally deferred to the optimization of the system. 
The paradigm draws heavily on the game theoretic underpinnings and is very much analogous to mechanism design. 
Traditionally, mechanism design maintains a relative indifference to the resulting behavioral dynamics so long as the desired end result is achieved -- what matters is the parameters the designer has placed on that end result, as well as a willful inattention to how those parameters are met. 
For example, autonomous vehicles might be made safe by first modeling the entire roadway and then creating behavioral incentives for all road users that would align their activities with the model specification.

Approaches taken in HIL Autonomy  developed from origins in control theory, and thus in contrast tend to begin with the underlying dynamics that define the domain (whether physical, informational, or social). 
As such, the dynamics serve to ground the model abstraction and are used with a formal verification procedure to guarantee that the system exhibits desired behaviors.
The dynamics themselves are treated as sacrosanct, either because they are natural and thus beyond designer control (like Maxwell’s laws describing electromagnetism) or protected by law or custom and thus outside the system specification \cite{dobbe2019toward}. 
To return to the above example, HIL Autonomy theorists would instead rely on deterministic physical constraints (e.g. physical laws of motion and following distances between vehicles) and human factors evaluations of the road environment to achieve safety through an analysis of \textit{reachability}: 
controlling to prevent the worst possible model state (i.e. crashes) by calculating all possible evolutions of an environment and likelihood of danger~\cite{bansal2017hamilton}.

Fair ML, however, relies predominantly on supervised learning within which domain features cannot be changed and which is not canonically represented in terms of feedback loops or evolution over time~\cite{corbett2018measure,madras2018learning}.
The technical focus is not on defining ends or protecting these dynamics but surgically intervening on them through “one-shot” \textit{classification} decisions in order to meet specified thresholds of fairness, accountability, or transparency.
ML is typically treated as a tool for precisely achieving institutional priorities that are already in place, like credit scoring~\cite{kamishima2011fairness}, school admissions~\cite{agarwal2018reductions}, and predictive policing~\cite{chouldechova2018frontiers}.
Correspondingly, Fair ML practitioners may be more concerned with accounting and assigning responsibility for crashes -- which may entail rigorous data collection on where and how autonomous vehicles are involved in crash events -- than minimizing their likelihood.

We can compare these approaches by appraising their assumptions about the interface between system and world.
In AI Safety, the model is meant to intervene on and structurally transform the dynamics of the domain in order to guarantee whatever ends the designer has specified.
In Fair ML, the model serves as a mere snapshot of the domain dynamics and is meant to administer the proper classification according to canonical metrics.
Meanwhile for HIL Autonomy, the model acts more as a hook into the ``natural'' dynamics of the domain, and is designed to respond to feedback in ways the designer can control.
In this way, each subfield is more or less intensely concerned with specification or optimization based on whether the goal of abstraction is to transform, manage, or redesign the dynamics of the domain. 
% These differences in how the problems are viewed and designed are reflected in the processes that fields use to arrive at agreement.

%% file: 3c_consensus.tex
\subsection{Consensus: What are methods for reaching agreement on risk and who is included in the process?}

Each subfield's orientation towards problems of interest is only partially determined by their distinctive tastes, values, and favored approaches to optimization and specification. 
The particulars of how subfields reach consensus on these topics is illustrative of further differences in how practitioners behave and develop technology. 
By contrasting the approaches to reaching consensus, it is possible to highlight the varying levels of deference to authority, legacy of regulatory structure, and different timescales of interest.

A common feature of the AI Safety community is \textit{forecasting} future AI capabilities against various time horizons based on expert opinions~\cite{grace2018will}. 
For example, it is common to forecast when autonomous vehicles will achieve full autonomy and when consumers will adopt said technology~\cite{bansal2017forecasting}.
This process favors the opinions of highly regarded researchers in AI and Computer Science. 
The problem of reaching consensus on the risks of AI development thus relies heavily on figures of authority who are largely members of technical fields without substantial social science training or affiliation. %\nol{added phrase here}. 
Interdisciplinary voices include vigorous discussions surrounding AI policy~\cite{brundage2020toward} and ethics~\cite{gabriel2020artificial} as they relate to forecasts of future technological capabilities, often based in long-term scenario planning.
This focus on far-off implications makes apparent the extreme timescales of interest to researchers in this subfield.
However, there is a corresponding focus on minimizing the uncertainty of predictions in order to balance these projections. 
The idea of exponential technological growth underlies the assumed importance of diligent, early investigation: progress will seem slower in the short-term and simultaneously be ever-accelerating in the long run.

Comparatively, the subfield of Fair ML is marked by its practice of interdisciplinary~\textit{debate}.
Positioned from existing frameworks of nondiscrimination law and theories of social justice, the community is comprised of academics in Law, Policy, and Science and Technology Studies in addition to Computer Science.
There is further an appeal to understand and improve the lived experiences of marginalized populations. 
In contrast to AI Safety's respect for authority coming from positions of power, the Fair ML community recognizes how individuals with limited power are often early indicators of harmful technology due to disproportionate exposure and compounding effects of overlapping technologies. 
The subfield thus has a clear focus on existing inequalities, while articulating concerns about how near-term deployment of tools will change social relations, such as automated scoring and surveillance~\cite{o2016weapons}. 
Fair ML champions the idea that the risks of AI take place on human timescales and have human impacts, so those affected should be part of the conversation. % regarding the scope of risk.

The subfield of HIL Autonomy rests on a legacy of public-private coordination and \textit{certification}. 
For example, the aerospace industry is characterized by collaborations between federal regulatory agencies (e.g. Federal Aviation Administration) and large companies (e.g. Boeing)~\cite{ilcewicz2005safety}.
% Engineering expertise and safety culture play an important role in developing regulation, but once created, these frameworks of regulation serve as the baseline specification without need for further debate.
Engineering expertise and safety culture play an important role in developing regulation, but once created, the resulting specifications do not usually require further debate.
There is a clear focus on broadly agreed upon notions of physical safety.
Even when metrics are contested, the process is better characterised as a re-evaluation than a deliberative or normative reflection.
Modern research on HIL Autonomy, such as algorithmic human robot interaction (HRI) looks to fields like psychology for tools to plan around human behavior, but isn't yet characterized by a strong culture of including voices of non-technical disciplines.
Furthermore, the subfield of HIL Autonomy is largely concerned with the instantaneous behaviors of physical systems, where rapid computation is necessary for preventing instabilities.
Rather than considering the accumulation of risk over years as technologies are developed, the main concern is on this minuscule timescale of harm caused by system instability.

%% file: 3d_failure.tex
\subsection{Failure: Is failure inevitable, how is it contained, and how is it measured?}
The way in which a domain handles failure is perhaps the most relevant to identifying regulatory interventions. 
Yet, the manner in which failure is abstracted varies across the three subfields we explore in this work. 
{The types of failure modes considered, who or what is meant to identify them, and the way through which they are addressed are all distinct. 
The modes of failure considered are essentially the counterparts of the value abstraction around which each field is oriented, while the manner in which failure may be identified and addressed is influenced by the optimization and consensus axes respectively. 
In this way, the conception of failure within each of the fields we discuss is very much a function of the abstractions discussed earlier in this section.} 
In this discussion it is useful to further decompose the failure abstraction into two components: \textit{robustness}, or how likely a system is to fail, and \textit{resiliency}, whether the system can fail safely. 
How AI Safety, Fair ML, and HIL Autonomy construe and emphasize these dimensions of failure provide valuable insights into their conception of the sociotechnical.

AI Safety is primarily concerned with the robustness of a system and the notion of safety by design. 
The aim is to develop \textit{provably beneficial} autonomous agents. 
If an agent is aligned to human values, it will act in the best interests of humanity, even when situations arise that had not been considered at design~\cite{russell2019human}. 
In this way, resilience is delegated to the optimization the agent performs post-deployment. 
Success, with regards to failure, is a system that adapts and avoids potential pitfalls. 
Accordingly, the conception of failure in AI Safety is predominantly focused on objective misalignment, or the mismatch between the objectives the system was designed to achieve and those it was optimized to achieve~\cite{bostrom2012superintelligent}. 
This conception of failure is captured in the classic Paperclip problem, in which an agent directed to efficiently produce paperclips ultimately destroys humanity on its singular path towards its objective. 
The stakes on which this dominant failure mode operate, motivate the off-switch intervention \cite{hadfield2016off}. 
In other words, if the system fails, the only recourse is to shut it down. 
In this sense, the notion of resilience is not explicitly optimized for in AI Safety. 

Fair ML on the other hand is oriented particularly around resilience. 
As a subfield, it operates on inherently opaque principles (e.g. fairness, equity, etc.), meaning that to some extent, system failure is inevitable. 
Though the values of the system are generally formalized and explicitly modeled, system failure is not. 
Rather, a safe interface between the the system and society is an iterative one. 
Resilience in Fair ML requires that the system be \textit{contestable}. 
Contestability is a discursive process that engenders system qualities like explainability and transparency~\cite{mulligan2019shaping}. 
If a system has the capacity to engender harm through this process or cannot accommodate proper iteration, it may be deemed an unfit application of ML~\cite{benjamin2016}. 
Notably, this notion of refusal has been considered in the context of facial recognition technology~\cite{acm2020statement}, a potentially high-stakes application in which safe iteration towards a more robust system is challenging. 

Finally, HIL Autonomy is oriented about both robustness and resiliency.
In other words, a successful system must be designed to operate safely, but must also be capable of failing safely. 
{Safety is often defined in terms of \textit{stability}, where all states of a system converge in a guaranteed manner to a desired value. 
% This draws on the notion of over-specification deriving from control systems engineering, often referred to as \textit{stability} or where all states of a system converge in a guaranteed manner to a desired value. 
Certification processes require that the likelihood of instability and failure remains small,
necessitating a complete accounting of the environment and explicit modelling of potential failure modes given a specified environment~\cite{reluplex2017}.
These stringent requirements often necessitate the practice of over-specification, in which the system is hierarchically decomposed so that failure modes are understood as the result of simultaneous failures in subsystems~\cite{brosgol2010178c}.
% Under this paradigm, it is necessary to completely account for the environment and explicitly model potential failure modes given a specified environment~\cite{reluplex2017}. 
% Under this paradigm, it is necessary to completely account for the environment and explicitly model potential failure modes given a specified environment~\cite{reluplex2017}. 
% Failure modes are hierarchically decomposed into distinct events that cause unintended state-space or dynamics~\cite{brosgol2010178c}
% The certification process 
% for reaching consensus within HIL Autonomy necessitates engagement on the part of the designer with the failure modes a system may encounter, often delineated as varieties of unintended state-space or dynamics~\cite{brosgol2010178c}. 
The probability that each subsystem fails must be suitably small, motivating an emphasis on robustness.
% For each of these failure modes, a system designer must then consider the substituent components contributing to the failure and ensure that even should the corresponding failure occur, each of them meets safety specifications. 
The probability that subsystems fail simultaneously must also be small, promoting redundancy and resilience in system design. 
In this way, the specification of safety in the context of HIL Autonomy guides system designers to hold explicit conceptions of both components of safety. }

%% file: 4_case.tex
\section{Case Study on Sociotechnical Inquiry: Consumer Drone Technology}
\label{sec:casestudy}

Intelligent consumer drones are beginning to interface with society in many ways: personal entertainment, commercial logistics, local security, and more~\cite{floreano2015science}.
In this section we study sociotechnical inquiry into consumer drone technology from the perspective of different AI subfields introduced in \sect{sec:subfields}.
The case study highlights the axes for comparison shown in~\tab{tab:tomstable}.
Individual consumer drones are gaining substantial capabilities by leveraging multiple advances in artificial intelligence, with computer vision, for example, enabling autonomous tracking and exploration in commercially available products~\cite{ackerman2018skydio,dji_official}.
% Individual consumer drones are gaining substantial capabilities by leveraging multiple advances in artificial intelligence, with computer vision, for example, enabling following targets and autonomously exploring complex environments in commercially available products~\cite{ackerman2018skydio,dji_official}.
Future iterations of these devices can take many forms of autonomy---from a security risk as \textit{slaughterbots}~\cite{russell2018you}, to virtual assistants that have more direct integration with the social~\cite{bachrach2017magic}, and to groups of agents working together to accomplish at-scale distributed tasks~\cite{vattapparamban2016drones}.

% \subsection{Human-in-the-loop Autonomy}
HIL Autonomy most directly integrates with the sociotechnical of consumer drones---the technology is studied from a lens of minimizing physical harm.
Drawing on its roots in control theory, HIL Autonomy considers consumer drones as an open problem of robust and agile control in the presence of human uncertainty. 
Broadly, the subfield optimizes the system dynamics by wrapping a bounding box around the {robot} itself, and the environment, which includes humans.
{At this point, a reachability problem can be formulated to provide numerical guarantees against human collision.}
% Such a problem can be formulated as reachability or other numerical guarantees against human collisions.
To determine which guarantees are necessary, designers might look to an official certification process for various behaviors.
The process of certification must be revisited, such as with recent updates to the Federal Aviation Administration's code permitting drone delivery at night and moving towards required pre-flight registration.
% HIL autonomy bounds the effects of the drones closely to that of its physical behavior, regardless of goal. 
HIL Autonomy arrives at consensus before and during deployment, primarily to consider the potential failure modes of the designed controllers (e.g. crashes due to algorithmic or mechanical error).

% \subsection{AI Safety}
As drones develop and gain more hierarchical and abstract versions of autonomy, AIs planning in the space of goals and intents may create concerns of mis-interpretting human values. 
AI Safety is concerned with long-term %behavior and
intelligence growth and the alignment between drones, drone swarms, and humans.
% AI Safety practitioners would perform optimizations away from danger of specific control commands to concern the long-term behavior of the system at large.
{AI Safety practitioners thus focus optimizations on the long-term behavior of the system at large rather than dangers of specific control commands.}
To arrive at agreement on the prevention of harm, system designers focus on forecasting outcomes and risks strictly before deployment.
% Such a practice is inherently limited to the accuracy of models of both the system, the environment, and the underlying intelligence. 
Specifically, AI Safety might look to permanently prevent cybersecurity risks around 
autonomous agents targetting humans with direct harm,
% actions of humans being the target of direct harm from autonomous agents, 
similar to the situation portrayed in \textit{Slaughterbots} with independent decisions~\cite{russell2018you}.
For an AI Safety practitioner, failure in this regard involves pervasive risk to society via numerous dangerous drones without well modeled intents.

% \subsection{Fair ML}
{Fair ML focuses on the interactions between drones and existing norms, both at the individual and societal levels.} %, separating away from specific dynamic control and constraining the intent of behavior.
The potential for harm arises due to models trained from frequently biased datasets with sometimes biased algorithms.
%the potential for failure and harm to undeserving individuals grows because all models are trained from frequently biased datasets with sometimes biased algorithms. 
For example, computer vision may be used to identify individuals for payments, monitoring, or other personally identifiable tasks. 
{Fair ML focuses on pinpointing specific forms of inequity---instead of optimizing a model parameter, such as validation accuracy, practitioners might search for solutions in the forms of better datasets and debiasing algorithms.
There may be a loop between
consensus and failure, where particular inequities motivate debates over solutions.} 
In contrast to other subfields, Fair ML might see some technological failures as indicative of an impossibility, and will refuse to accept certain tools (such as for surveillance). 
To arrive at consensus after an adverse event, stakeholders are ideally included in the conversation, a practice of great value to development of the sociotechnical.

Both AI Safety and Fair ML take the broader lens and consider how the behavior of a drone, or a swarm of drones, can re-define sets of interactions within a given society.
% Beyond individual and group interactions, only Fair ML and AI Safety can also take a broader lens where the behavior of a drone, or a swarm of drones, can re-define sets of interactions within a given society. 
The potential for surveillance and control with at-scale adoption on intelligent drones poses critique from the subfields in different ways.
% This at-scale-application of drones provides a second entry point into consideration of the sociotechnical: how such systems can augment power structures, especially when profiling on demographic properties.
It provides a second entry point into consideration of the sociotechnical: how such systems can augment power structures or perpetuate demographic profiling.
The Fair ML community may focus on mitigating harms of individual interactions, but when harm is not controllable, they opt for a flat refusal of technological adoption.
AI Safety would view the at-scale swarm problem relatively similarly to the epistemic alignment of individual drones.
% 's view of this new swarm problem would be relatively similar to the individual drone as epistemic alignment.
{These differences highlight how current problems in HIL Autonomy are less likely to change dramatically in scope, as they are focused on improving the capabilities of individual agents.} 
On the other hand, as increasingly intelligent behaviors become possible, the views of AI Safety and Fair ML practitioners are more likely to develop at the same pace of technological development.
As with any technology, the future path of consumer drone development and adoption is not certain, so regularly re-evaluating pathways of future use is crucial.

%% file: 5_concl.tex
\section{Conclusion}
Given the pace and breadth of the field, constructive tools for sociotechnical inquiry into AI research and technological development are needed.
In this paper, we have presented a framework for sociotechnical inquiry along axes of \textit{value}, \textit{optimization}, \textit{consensus}, and \textit{failure} as an initial toolbox to gauge a research domain's relationship with society.
The subfields of AI Safety, Fair ML, and HIL Autonomy do not represent a total or exhaustive landscape of social interfaces of AI development, but those fields that have matured recently towards their own lens of sociotechnics. 
Future study in other wide directions of AI research such as cybersecurity, ML in healthcare, and more are strongly warranted as directions for future work.
We hope that this paper helps nurture a community of engaged scholars examining AI's possible interfaces with society, and encourages further lexicons that facilitate translations between technological and social development.